\documentclass[letter, 10 pt, conference]{ieeeconf}  % Comment this line out
\IEEEoverridecommandlockouts%                              % This command is only
\usepackage{graphicx}
\usepackage{epsfig} % for postscript graphics files
\usepackage{url}
\usepackage[flushleft]{threeparttable}
\usepackage{cite}
\usepackage{rccol}
\usepackage{booktabs}
\usepackage{tabulary}
\usepackage{times} % assumes new font selection scheme installed  <<<< je ne sais pas si on doit l'utiliser ?
\usepackage{subfigure}

\usepackage[tophrase=--]{siunitx}
\usepackage{epstopdf}
\usepackage{todonotes}

\newcommand{\otoprule}{\midrule[\heavyrulewidth]}

\title{\LARGE \bf Using Scalp Electrical Biosignals to Control an Object by Concentration and Relaxation Tasks: Design and Evaluation
}

\author{Laurent George, Fabien Lotte, Raquel Viciana Abad, and Anatole L\'ecuyer% <-this % stops a space
\thanks{L. George, F. Lotte, and A.L\'ecuyer are with INRIA, France,  \{laurent.f.george, fabien.lotte, anatole.lecuyer\}@inria.fr. R. Viciana Abad is with Department of Telecommunication Engineering, University of Ja\'en, Spain, rviciana@ujaen.es. 
}
}

\begin{document}

\maketitle
\thispagestyle{empty}
\pagestyle{empty}

\begin{abstract}
  In this paper we explore the use of electrical biosignals  measured on scalp and corresponding to mental relaxation and concentration tasks in order to control an object in a video game. To evaluate the requirements of such a system in terms of sensors and signal processing we compare two designs. The first one uses only one scalp electroencephalographic (EEG) electrode and the power in the alpha frequency band. The second one uses sixteen scalp EEG electrodes and machine-learning methods. The role of muscular activity is also evaluated using five electrodes positioned on the face and the neck.
  Results show that the first design enabled 70\% of the participants to successfully control the game, whereas 100\% of the participants managed to do it with the second design based on machine learning. Subjective questionnaires confirm these results: users globally felt to have control in both designs, with an increased feeling of control in the second one.  
 Offline analysis of face and neck muscle activity shows that this activity could also be used to distinguish between relaxation and concentration tasks. Results suggest that the combination of muscular and brain activity could improve performance of this kind of system. They also suggest that muscular activity has probably been recorded by EEG electrodes.
 % The study was conducted with 10 subjects. Participants were asked to control the altitude of a spaceship displayed on a screen by voluntarily doing a mental relaxation task to make the spaceship go down and a concentration task to make the spaceship go up. 
 %Results of this experiment are surprisingly good and suggest that both setups seem well suited for interaction with simple video games. The first design enabled 70\% of the participants to successfully control the spaceship, whereas 100\% of the subjects managed to do it with the second setup based on more complex machine learning. Furthermore, offline analysis of face and neck muscle activity highlights that this activity could also be used as a discriminator between relaxation and concentration task. Thus the combination of electromyography (EMG) and EEG %Questionnaires also revealed that tiredness started to occur after 9 minutes of control on average. This suggests that both design are well suited for short interaction in video games.  
\end{abstract}

\section{INTRODUCTION}
 %A Brain-Computer Interface (BCI) is a communication system which uses brain activity as an input channel~\cite{wolpaw2002brain}. Different techniques exist to provide brain signals to BCI such as Electroencephalography (EEG), which measures the electrical activity generated by the brain. So far, the main brain signals that have been employed to drive a BCI are Evoked potentials and brain activity related to motor imagery. 
 In recent years, researches have been conducted to use electrical brain activity for interacting with video games and virtual environments~\cite{lecuyer2008,Plass-OudeBos2010,blankertz2010berlin}.
 In this context, researchers have started to explore signals correlated to concentration and relaxation states~\cite{brainball2000,lin2006,muehl2010}. Several electroencephalographic (EEG) markers have indeed been identified as correlated with mental relaxation, task engagement, attention and mental workload. For example, alpha rhythm (\SIrange{8}{12}{\hertz}) is known to be attenuated or blocked by mental activity and attention~\cite{niedermeyer05}.   
Other rhythmic activities such as beta and theta rhythms have been investigated to quantify mental relaxation~\cite{lin2006} and attention level~\cite{Pope1995}. Hamadicharef \textit{et al.}~\cite{hamadicharef2009} have also assessed the level of attention through EEG, but employing machine learning methods to learn the best frequency and spatial filters to use. This approach reveals better classification accuracy than using only alpha, beta and theta bands. 
 More recently M\"uhl \textit{et al.}\ proposed to use alpha activity as a passive Brain Computer Interface (BCI) modality~\cite{muehl2010}. Alpha activity was used in an implicit way to adapt the game content. Enemies speed was increased when the alpha activity was low (which was expected to be correlated with a low relaxation state). Unfortunately, no significant differences for game experience and for user's alpha power modification were found. 
Electrical activity has also been used in an explicit way. In the game BrainBall~\cite{brainball2000}~(later adapted into a commercial game called Mindball\footnote{\url{http://www.mindball.se/product.html}}) players must be relaxed to move a ball away. A ratio between beta and alpha activity measured with forehead electrodes was used to control the ball position. 
Matel Inc.\ also proposed the MindFlex games\footnote{\url{http://mindflexgames.com/}} in which the player has to concentrate to control physical objects. However, we do not have information about the electrical signal and the signal processing used in these games. 
Despite these promising first results, there is still a number of important scientific questions about such systems that need to be answered. In particular:
\begin{itemize}
  \item Quantitative and in depth performance analysis are scarce in the litterature;
%\item There is a lack of rigorous evaluations of performances; 
  \item Due to a lack of comparative study, the best electrodes~(number and placement) and signal processing to be used are still not well known; %%% TODO 
%\item The design requirements in terms of sensors to be used (electrodes number and placement), signal processing (frequency band, etc.) are not well understood;
\item There is still a debate about whether such systems are pure EEG systems, independent of Electromyogram (EMG) activity, or if EMG plays an important role~\cite{loosermentalblock}. 
\end{itemize}
Thus, in this paper we have conducted a study in which users had to perform mental concentration and relaxation tasks to control an object in a video game, based on EEG recordings, which aims at addressing the aforementioned questions.

The remainder of this paper is organized as follows. In Section~\ref{sec:method}, the application developed and the experimental procedure are described. In Section~\ref{sec:results}, results are presented and discussed. Finally, the main conclusions are summarized in Section~\ref{sec:conclusion}.

%The feedback provided to the user is continuous.
\section{METHOD}
\label{sec:method}
%In the present study, an experiment in which users had to do mental concentration and relaxation tasks to control an object in a video game was conducted. This section describes the game design. 
\subsection{Application}
\begin{figure}
  \centering{
\subfigure{
		\includegraphics[width=0.45\linewidth]{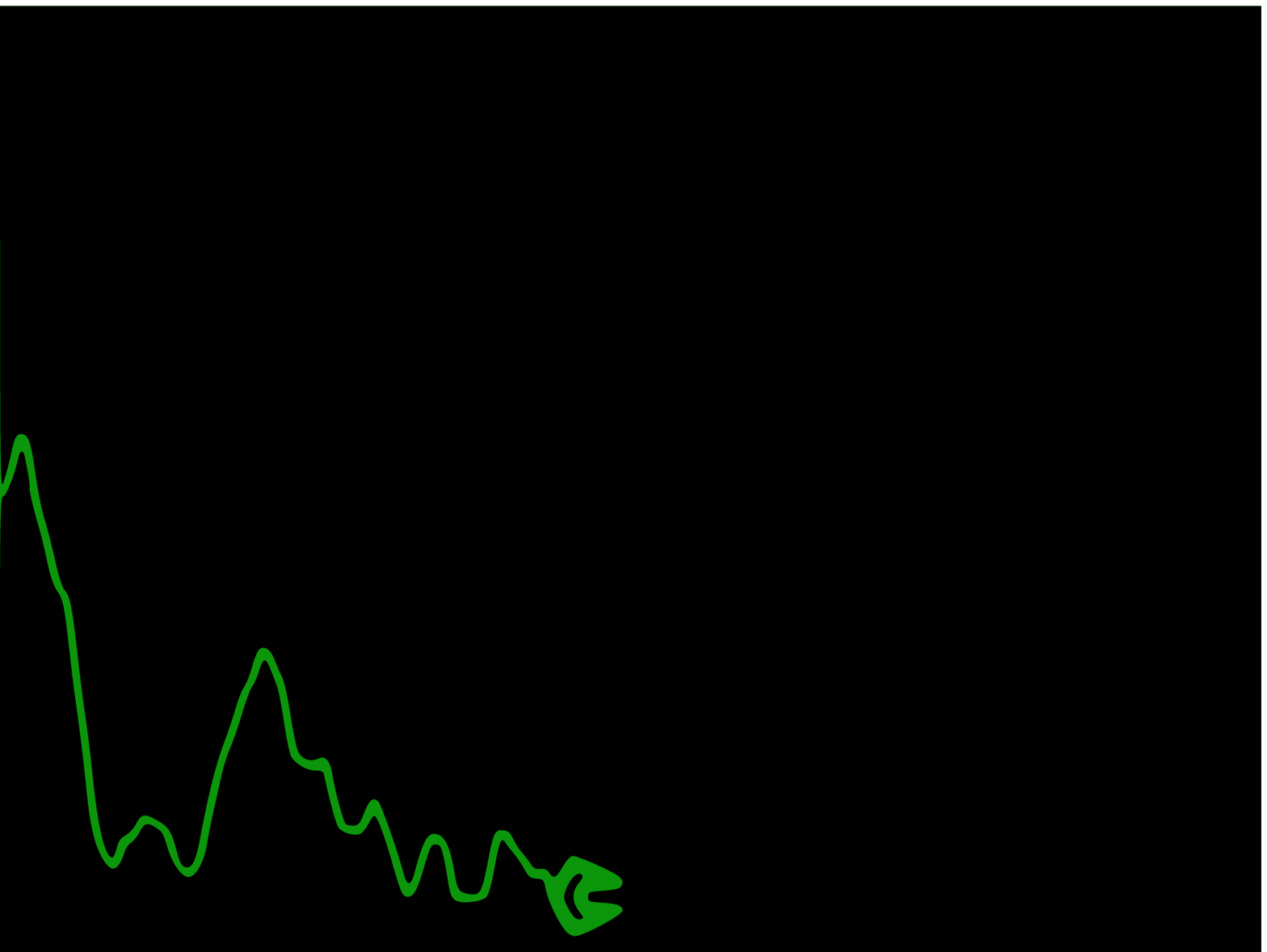}
	}
	\subfigure{
	\includegraphics[width=0.45\linewidth]{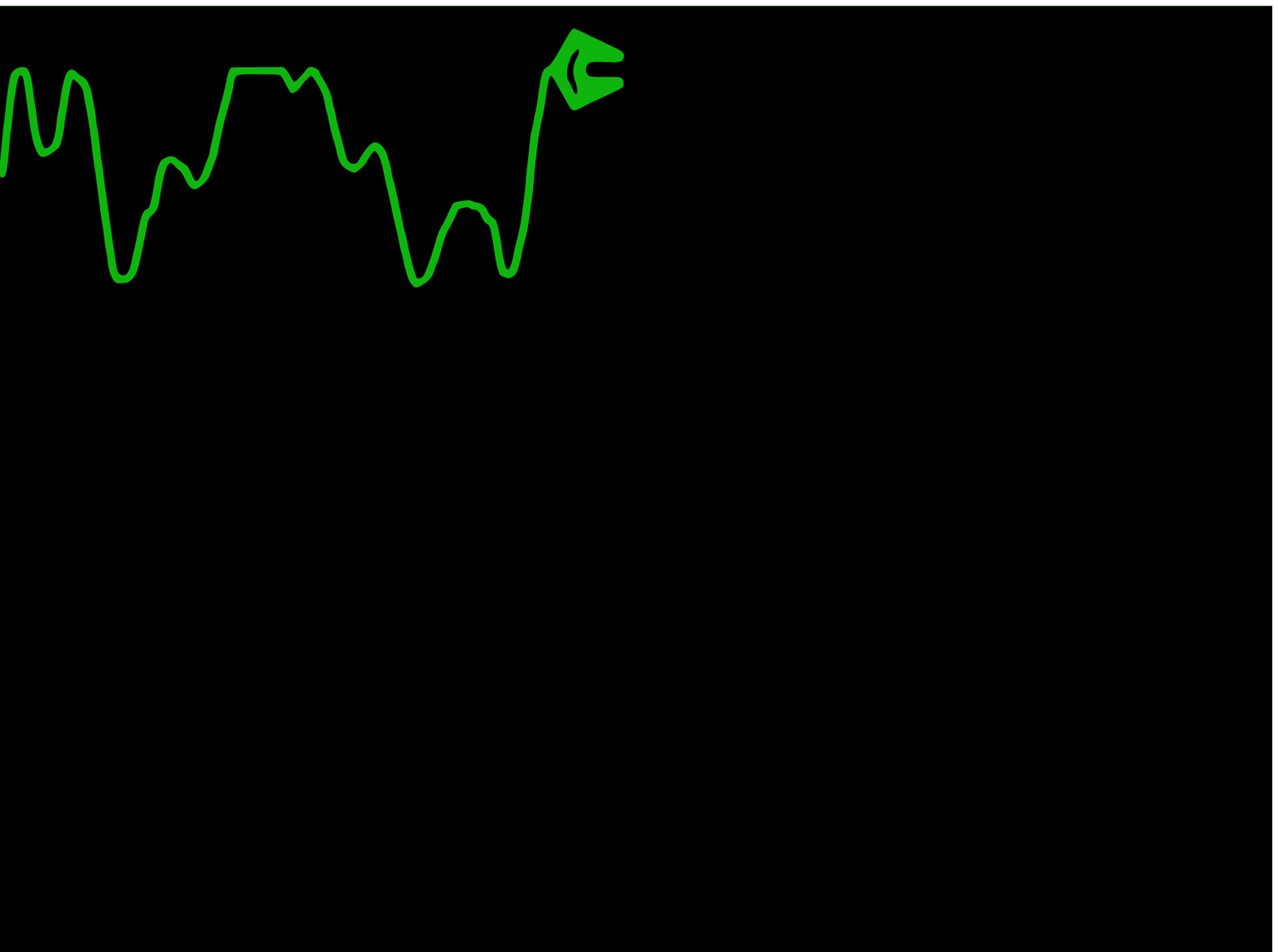}
	}
	\caption{The video game application with feedback during the two different phases (Go Down on the left, Go Up on the right).\label{fig:screenshot}}
	}
\vspace{-0.30cm}
\end{figure}

The goal of the game was to control the altitude of a spaceship displayed on the screen. Two kinds of instructions were provided to the user: ``Go Up'' and ``Go Down''. When ``Go Up'' instruction was displayed, the user had to try to make the spaceship go up and stay at a high altitude during \SI{35}{\second}. The opposite behavior was expected for the ``Go Down'' instruction, by making it go down. The spaceship altitude depended on the user's level of concentration or relaxation as computed by the system (see section~\ref{sect:signal}). The maximum and minimum altitudes were respectively $1$ and $-1$. A green line that symbolizes the smoke of the spaceship was used to provide visual feedback to the participants. Figure~\ref{fig:screenshot} displays a screenshot of the application during typical ``Go Down'' and ``Go Up'' phases. 
\subsection{Procedure} %% experimental design ? 
The experiment was carried out in an office room. Participants sat in front of a $24^{\prime\prime}$ LCD screen device. The distance to the screen was \SI{90}{\cm}. Before the experiment began, all participants were informed about the aim of the study and the relation between the spaceship position and their mental state; and they filled out a consent form. To make the spaceship go up, they were suggested to follow one of the two strategies: \textit{``focus your attention on the spaceship altitude''} and \textit{``imagine sending mental orders to the spaceship such as Go up!''}.
To make the spaceship go down, they were suggested to follow one of the two strategies: \textit{``do not focus your attention on the spaceship and its altitude''} and  \textit{``focus your attention on your respiration patterns to relax''}. Participants were instructed to look at the display screen during both conditions. They were also informed that the recorded signals are sensitive to artifacts (like body movement, teeth clenching), for these reason they should move as little as possible throughout the experiment. 

The experiment consisted of two parts (Part~A and Part~B) corresponding to two setups (a version based on alpha activity and a version based on machine learning methods, respectively). To assess differences between the two setups, the experiment had a within-subjects design. Thus all the participants performed each part of the experiment on a different day, separated by less than a week. To eliminate possible order effects between versions, half of the participants started with the first version (group A did Part A first) and the other half with the second version (group B did Part B first).
% peut etre mettre cette phrase au debut de la partie questionnaire..

%One aim of the study was to check if alpha rhythm measurement with an unique electrode and a simple processing chain was sufficient to control the object. We also wanted to compare this simple version to a more sophisticated one where 16 electrodes were used. For these reasons two different versions of the game were developed, and a within-subjects design was used to compare them. All participants played both versions of the game. To eliminate bias due to play's order between versions, half of the participants started with the first version when the other half started with the second. 

%The experiment was divided into two parts (part~A and part~B) corresponding to both versions of the game (respectively version based on alpha activity and version based machine learning methods). Each part occurred on a different day, separated by less than a week. All participants did both parts. They were randomly separated into two groups~: group A and group B. Group A started with Part~A before doing Part~B (resp.\ group B started with Part~B before Part~A).
Each part of the experiment consisted of six sessions. Each session was divided into three blocks. Each block had two phases: one ``Go up'' and one ``Go Down'', starting with a randomized order. Phase duration was \SI{35}{\second}. At the start of each game, the instructions were displayed at the top of the application screen for \SI{5}{\second} (``Go up'' or ``Go down''). The spaceship was displayed and its altitude was directly mapped to the control signal (see section~\ref{sect:signal}). The blocks were separated by a break of \SI{10}{\second}, during this time the screen was blanked. Each session was separated by a break of less than \SI{2}{\minute}. This time was used to run the machine learning process between sessions (see section~\ref{sect:signal}).
%Each part comprised 6 sessions. Each session was divided into three blocks. A block corresponds to a Go up phase and a Go down phase (in a randomized order). At the start of a phase (Go Up or Go Down), instructions were written at the top of the application screen for five seconds. The spaceship was displayed and its altitude was updated based on the control signal. Blocks were separated by a break of 10 seconds, during this time the screen was blanked.  Phase duration was 35 seconds. Each session was separated by a break of less than 2 minutes. This time was used to run the machine learning process between session (see section~\ref{sect:signal}).

\subsection{Participants}
Ten participants (8 men, 2 women) aged from 22 to 34 ($\mu=26.2$, $\sigma=4$) took part in the experiment. Subjects were randomly assigned to the two different groups A and B.
%% pas pour ce papier ...\todo{ON laisse ? car on n'en parle pas apres} Four participants (3,6,7,8) had already used a BCI system, while the remaining ones were naive to BCI. 

\subsection{Electrodes montage}
Two sets of electrodes were used to record data throughout the experiment. The first set of electrodes (EEG set) was used to record EEG activity. The purpose of the second set (EMG set) was to record EMG activity on the face and the neck. 
%\footnote{\url{http://www.gtec.at}}
A Gtec UsbAmp was used to acquire EEG data at a sampling rate of \SI{512}{\hertz} (EEG set). This device enabled to use 16 electrodes according to the 10-20 international system. It was used to record EEG activity related to mental relaxation and concentration tasks. The electrodes used were: Fp1, Fpz, Fp2, F3, Fz, F4, T7, C3, Cz, C4, T8, P3, Pz, P4, O1 and O2.  Electrodes Fp1, Fp2, Pz were chosen accordingly to BCI literature~\cite{lin2006,muehl2010}. The remaining 11 electrodes were placed on a grid pattern to cover an important surface of the scalp. A ground electrode (located on the left ear) and a reference electrode (located at AFz position) were also used.

%\footnote{\url{http://www.mindmedia.nl}}
To record electrophysiological activity on the user face and neck a MindMedia NeXus 32b device was used (EMG set). Four electrodes were placed on the face (above and below each eye), and one electrode was placed on the neck. A ground electrode (located on the nose) and a reference electrode (located on the forehead) were also used with this setup. These electrodes were not used for online control of the game but only for offline EMG analysis.

\subsection{Signal Processing}
\label{sect:signal}
Signal acquisition and online processing were conducted using the open-source software OpenViBE~\cite{openvibe}. Offline processing was done using Matlab. 
Two control signals were used during the online experiment. The first one was computed using only \textbf{1 scalp EEG electrode} and power in the alpha frequency band (CS1). The second one was acquired from \textbf{16 scalp EEG electrodes} and processed by applying machine learning methods (CS2). 
%alpha Pz % sinon y a ce livre aussi : Fisch and Spehlmann's EEG Primer 
%% relire seconde version bacteria hunt pour valider le choix alpha Pz.. me semble qu'ils citaient un autre papier

The first \textbf{control signal (CS1)} corresponds to the opposite of the power of the user's alpha frequency band at Pz location. Indeed alpha activity is known to increase when the user is relaxed~\cite{niedermeyer05}. As this activity is mainly found in the posterior half of the head~\cite{niedermeyer05}, the electrode was placed on Pz position. This activity was quantified by a band power technique~\cite{pfurt1999bandpower}. Band power was computed 16 times per second on a moving window of \SI{1}{\second}. The alpha band selection was based on the method of individually adjusted bandwidths (IBFW) proposed by Doppelmayr \textit{et al.}~\cite{doppelmayr1998}. To determine the individual dominant alpha frequency of each participant, we used a \SI{30}{\second} recording of the EEG data with participants closing their eyes. Then, the spectrum of the signal was computed and the frequency (in \SIrange{6}{20}{\hertz}) that shows the maximum power was chosen. A fixed width of \SI{4}{\hertz} around this value was then used as the subject's alpha frequency band. A calibration phase with no feedback was used to scale the control signal between $-1$ and $1$. The calibration phase corresponded to a \SI{30}{\second} relaxation task and \SI{30}{\second} concentration task, both without any feedback. 

To compute the second \textbf{control signal (CS2)}, a technique similar to the one proposed in~\cite{hamadicharef2009} was used. All EEG signals were passed through a bank of \SI{4}{\hertz} bandwidth filters centered on all the frequencies between \SI{2}{\hertz} and \SI{30}{\hertz}. A common spatial pattern (CSP) method~\cite{blankertzcsp} was then used to compute spatial filters for each of them. Maximum relevance feature selection was used to select the five most relevant couples of frequency band and spatial filter~\cite{mutualinforef}. The linear model that best matched features (computed using the selected band pass filters and spatial filters) and both classes (Up and Down) was then determined using linear regression on the training set. The linear equation, selected band filters, and selected spatial filters were then used online. The learning process was run between each session using all data recorded during the previous sessions of the day.

In part~A, CS1 was used during the whole experiment (altitude of the spaceship was mapped to CS1). In part~B, CS1 was used for the first session to provide feedback to the user when collecting training data for the subsequent CS2 calibration (machine learning); CS2 was used during all the remaining sessions.
\section{RESULTS}
\label{sec:results}
\subsection{Subjective questionnaire}
\begin{table}
  \begin{center}
  \begin{threeparttable}[c]
  \caption{Questionnaire marks reported (mean$\pm$std)\label{table:question_histo}.}
  \begin{tabulary}{\columnwidth}{Lcc}
	\toprule%
	Question & Part A (CS1) & Part B (CS2)  \\
	\otoprule%
	%1.\ Control felt during the ``Go Up'' condition  \scriptsize{(1:~never, 7:~always)} &  $4.4\pm1.51$   &  $4.9\pm1.29$    \\
	Did you feel in control of the spaceship during ``Go Up''? ~~  \scriptsize{ (1:~never, 7:~always)} &  $4.4\pm1.51$   &  $4.9\pm1.29$    \\

	Did you feel in control of the spaceship during ``Go Down''?\scriptsize{ (1:~never, 7:~always)} &  $4.4\pm1.07$   &  $5.45\pm1.26$    \\
	
	Did you feel fatigue related to the interaction?\scriptsize{ (1:~very, 7:~not at all)}  & $4.3\pm1.25$ & $4.9\pm1.20$ \\
	How fun was the game? ~~~ \scriptsize{ (1:~boring, 7:~enjoying)}   &   $3.8\pm1.67$   & $4.2\pm1.48$\\
	\bottomrule%
  \end{tabulary}
  \begin{tablenotes}
  \item The first column corresponds to the use of alpha band power technique with one electrode Pz (CS1) during part~A. The second column corresponds to the use of 16 electrodes and  machine-learning method (CS2) during part~B.% Values are presented as mean$\pm$std.
\end{tablenotes}
	\end{threeparttable}
	\vspace{-0.4cm}
\end{center}
\end{table}
After each part of the experiment, participants were asked to fill in a questionnaire. Table~\ref{table:question_histo} provides the average marks they gave for each question. Results showed that participants felt that they had control over the spaceship in both parts. Indeed, the reported sensation of control (average marks given for question 1 and 2) was on average $4.4$ for Part~A and $5.17$ for Part~B on a scale between 1 (never) and 7 (always). 
During Part~A, there was no significant difference between the ``Go up'' and ``Go down'' conditions (Wilcoxon paired test $p=.95$). Concerning part B, there seemed to be a difference between ``Go up'' and ``Go down'' condition, but it was not significant (Wilcoxon paired test $p=.35$). Concerning the influence of the control signal, a significantly better feeling of control was observed in Part~B (Wilcoxon paired test $p=.02$). 
%Subject 8 is the only one that reported a smaller feeling of control for part~B than for part~A. 

Questionnaire results about \textit{tiredness} seemed to indicate that participants felt the experiment relatively tiring. Part~B was reported to be less tiring, although this difference was not statistically significant (Wilcoxon paired test $p=.20$). Participants were also asked to report when they started to feel fatigue during the experiment. Fatigue seems to start after the middle of the experiments (\SI{54}{\percent} completed for Part A, and \SI{62}{\percent} completed for Part B). In the middle of the experiment, participants have already completed \SI{9}{\minute} of active control (``Go up'' and ``Go down''). 

Concerning the \textit{fun}, it seems that participants felt an average enjoyment during the game. This could be due to the simplicity of the interface and the absence of rewards during the game. No significant difference was found between the two parts (Wilcoxon paired test $p=.28$).

\subsection{Online Performances}

\begin{table}[b]
  \begin{threeparttable}
	\caption{Performances expressed as difference between spaceship altitude in each condition (${Z}_{up} - {Z}_{down}$).\label{tab:perfonline} }
  \centering{
  \begin{tabulary}{\columnwidth}{|L|LL|LL|}
	\hline
	\multicolumn{1}{|c|}{} & \multicolumn{2}{c|}{Online} & \multicolumn{2}{c|}{Offline} \\
	\multicolumn{1}{|c|}{N.} & \multicolumn{1}{c}{Part A (CS1)}      & \multicolumn{1}{c|}{Part B (CS2)}      & \multicolumn{1}{c}{Part B (CS3)}     & \multicolumn{1}{c|}{Part B (CS4)}\\
	\multicolumn{1}{|c|}{} & \multicolumn{1}{c}{EEG}   & \multicolumn{1}{c|}{EEG}   & \multicolumn{1}{c}{EMG}   & \multicolumn{1}{c|}{EEG + EMG} \\
	%& \multicolumn{1}{c}{alpha Pz} & \multicolumn{1}{c|}{CSP} & \multicolumn{1}{c}{CSP} & \multicolumn{1}{c}{CSP}\\
		\hline
		1  & \textbf{-0.07}~(0.24)                 & \textbf{0.22}~(0.30)$^{\ast\ast}$     & \textbf{0.24}~(0.16)$^{\ast\ast\ast}$ & \textbf{0.42}~(0.34)$^{\ast\ast\ast}$ \\
		2  & \textbf{0.18}~(0.28)$^{\ast}$         & \textbf{1.03}~(0.21)$^{\ast\ast\ast}$ & \textbf{0.55}~(0.22)$^{\ast\ast\ast}$ & \textbf{1.32}~(0.16)$^{\ast\ast\ast}$ \\
		3  & \textbf{0.30}~(0.25)$^{\ast\ast}$     & \textbf{0.95}~(0.28)$^{\ast\ast\ast}$ & \textbf{0.95}~(0.32)$^{\ast\ast\ast}$ & \textbf{1.14}~(0.29)$^{\ast\ast\ast}$ \\
		4  & \textbf{0.32}~(0.16)$^{\ast\ast\ast}$ & \textbf{0.42}~(0.25)$^{\ast\ast\ast}$ & \textbf{0.37}~(0.14)$^{\ast\ast\ast}$ & \textbf{0.56}~(0.35)$^{\ast\ast\ast}$ \\
		5  & \textbf{-0.05}~(0.13)                 & \textbf{0.24}~(0.31)$^{\ast}$         & \textbf{0.30}~(0.17)$^{\ast\ast\ast}$ & \textbf{0.79}~(0.29)$^{\ast\ast\ast}$ \\
		6  & \textbf{0.22}~(0.11)$^{\ast\ast\ast}$ & \textbf{0.91}~(0.16)$^{\ast\ast\ast}$ & \textbf{0.17}~(0.23)$^{\ast}$         & \textbf{0.92}~(0.24)$^{\ast\ast\ast}$ \\
		7  & \textbf{0.05}~(0.26)                  & \textbf{1.02}~(0.39)$^{\ast\ast\ast}$ & \textbf{0.35}~(0.19)$^{\ast\ast\ast}$ & \textbf{1.06}~(0.42)$^{\ast\ast\ast}$ \\
		8  & \textbf{0.31}~(0.13)$^{\ast\ast\ast}$ & \textbf{0.35}~(0.29)$^{\ast\ast}$     & \textbf{0.17}~(0.20)$^{\ast\ast}$      & \textbf{0.44}~(0.38)$^{\ast\ast\ast}$ \\
		9  & \textbf{0.20}~(0.14)$^{\ast\ast}$     & \textbf{0.56}~(0.29)$^{\ast\ast\ast}$ & \textbf{0.15}~(0.13)$^{\ast\ast\ast}$ & \textbf{0.51}~(0.25)$^{\ast\ast\ast}$ \\
		10 & \textbf{0.13}~(0.19)$^{\ast}$         & \textbf{0.57}~(0.33)$^{\ast\ast\ast}$ & \textbf{0.31}~(0.22)$^{\ast\ast\ast}$ & \textbf{0.56}~(0.32)$^{\ast\ast\ast}$ \\
		\hline
\textbf{$\mu$} & {\textbf{0.16}} & {\textbf{0.63}} & {\textbf{0.36}} & \multicolumn{1}{l|}{\textbf{0.77}} \\
\hline
\end{tabulary}
}
\begin{tablenotes}
\item The first column corresponds to the use of alpha band power with electrode Pz (CS1). The second column corresponds to the use of machine-learning methods with 16 EEG electrodes (CS2). Offline performances using EMG set are presented in column 3. Offline performances using the combination of EEG and EMG sets are provided in column 4. Last line provides the average altitude difference over participants. Values are presented as \textbf{mean}~(std); asterisks represent the significance of the difference ($^{\ast}$: significant at $p<.05$; $^{\ast\ast}$: significant at $p<.01$; $^{\ast\ast\ast}$: significant at $p<.001$)
\end{tablenotes}
\end{threeparttable}
\end{table}

The first \SI{5}{\second} of each phase were ignored for the analysis of the online performances (it corresponded to the instruction reading time). We assessed performances by computing the average difference between the altitude in the two conditions (``Go up'', ``Go down'') over each block. Indeed, a higher altitude in the ``Go up'' condition could be interpreted as a success. The online performance indexes are provided in Table~\ref{tab:perfonline}.
%% on ne parle pas de la première session TODO TODO TODO
Spearmann correlation test between performance index and participant's control marks indicates that the difference between conditions seems to be correlated to reported marks (Part~A $r=.56$, $p=.09$; Part~B $r=.72$, $p=.02$).

The computation of a Student t-test, between conditions ``Go up'' and ``Go down'' in each block shows that there is a significant difference between the two conditions for seven out of the ten participants with CS1 (Part A), and for all the participants with CS2 (Part B) (asterisks in Table~\ref{tab:perfonline}). When the difference is significant it is also always positive: the spaceship altitude in the ``Go up'' condition is higher than during the ``Go Down'' condition. 
Regarding to the influence of the control signal it appears that the use of CS2 with 16 EEG electrodes provided better overall performances compared to CS1 (Wilcoxon paired test $p=.002$). It highlights the need of machine learning and subject specific design and confirms results in~\cite{hamadicharef2009}.

\begin{figure}[t]
	\centering{
	\includegraphics[width=0.75\columnwidth]{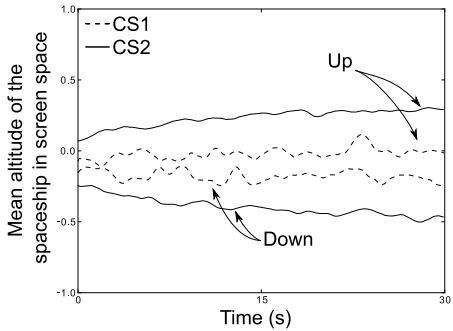} 
	\caption{Mean altitude of the spaceship on screen as function of different conditions (``Go-Up'' vs. ``Down'' and CS1 vs. CS2). Altitude was averaged over all blocks and all participants. Dotted line represents altitude when alpha frequency band at Pz electrode was used (CS1). Solid line represents altitude when machine-learning method was used with 16 electrodes (CS2).\label{fig:evolution_session}} 
	}
	\vspace{-0.4cm}
\end{figure}
The analysis of performances evolution over time is shown in Figure~\ref{fig:evolution_session}. It provides the average evolution of altitude for both conditions and both control signals. We observed an increase of performance over the course phase duration with CS2. This suggests that users need time to get into the task.
%%% c'est pas vraiment ca .. mais bon
\subsection{Role of EMG}
Co-occurrence of muscular activity is known to result in contamination of EEG by Electromyogram~(EMG)~\cite{Goncharova2003}. Therefore an evaluation of EMG activity during the interaction is necessary. For this purpose signals recorded during Part~B using face and neck electrodes~(EMG set) were analyzed. We computed a control signal (CS3) based on EMG~set employing the same machine learning methods as the one used for CS2 (see section~\ref{sect:signal}). The usage of EMG~set led to a significant difference between the ``Go Up'' and ``Go Down'' conditions for all the participants (Table~\ref{tab:perfonline}). This means that the recorded information in facial and neck electrodes can be used to discriminate between the two conditions although the use of EEG set (CS2) provided better performances (Wilcoxon paired test $p=.033$). One interpretation of these results is that EMG set measured some EEG activity, which is enough to discriminate between tasks (indeed electrodes above the eyes are able to measure some frontal EEG activity). A more likely interpretation is that EMG set measured EMG activity, which provides information to discriminate between tasks. This would mean that a discriminant muscular activity occurred when users try to do the two mental tasks, however they were asked to try to avoid muscular activity during the experiment. This second interpretation would also imply that the EEG set measured a part of this discriminant muscular activity.

To evaluate the combination of EEG and EMG sets in terms of performance the computation of a control signal (CS4) based on the three most relevant EEG features and the two most relevant EMG features was done (to have 5 features as with EEG or EMG alone). The combination of EMG and EEG allowed to increase performances in terms of difference between conditions compared to CS2 (Wilcoxon paired test $p=.019$). This result may indicate that there are relevant information in the two sets of electrodes that are not redundant between the sets. Future work should be done to quantify the part of information in each set and to qualify the origin of this information (muscular versus brain activity). 
\section{CONCLUSION}
\label{sec:conclusion}
 %Results of this experiment are surprisingly good and suggest that both setups seem well suited for interaction with simple video games. The first design enabled 70\% of the participants to successfully control the spaceship, whereas 100\% of the subjects managed to do it with the second setup based on more complex machine learning. Furthermore, offline analysis of face and neck muscle activity highlights that this activity could also be used as a discriminator between relaxation and concentration task. Thus the combination of electromyography (EMG) and EEG %Questionnaires also revealed that tiredness started to occur after 9 minutes of control on average. This suggests that both design are well suited for short interaction in video games.  
In this paper, we have explored the use of two different setups based on scalp electrical biosignals corresponding to mental relaxation and concentration tasks to control a video game. Results of this experiment are surprisingly good and suggest that both setups seem well suited for interaction with simple video games.  Analysis of signal from face and neck electrodes (EMG set) suggests also that there is relevant information in muscular activity correlated to the two different tasks and that the combination of EEG and EMG electrodes in a hybrid BCI~\cite{hybridbci} paradigm could improve performance. They also suggest that the current BCI approaches using EEG to detect concentration and relaxation tasks (e.g.~\cite{brainball2000}) may not be pure BCI as they may be involuntarily using EMG signals recorded with EEG electrodes. 
%Determining to which extent concentration and relaxation tasks can be decoded independently by EEG and EMG should be addressed in future works.
%The first design enabled 70\% of the participants to successfully control the spaceship, whereas 100\% of the subjects managed to do it with the second setup based on more complex machine learning.
%~\\
%~\\
%\footnotesize{
\section*{ACKNOWLEDGMENTS}
\noindent This work was supported by the French National Research Agency within the OpenViBE2 project and grant ANR-09-CORD-017.
%}

\bibliographystyle{IEEEtran}
%\small{
\bibliography{IEEEabrv,embcbib}
%}

\end{document}